\def\rfr#1{eq. (\ref{#1})}

\def\dert#1#2{\frac{{{d}}{#1}}{{{d}}{#2}}}              % derivate parziali e totali prima e seconda

\def\bb#1#2#3{\bibitem[\protect\citeauthoryear{#1}{#2}]{#3}}
\def\eqi{\begin{equation}}
\def\eqf{\end{equation}}
\def\eqia{\begin{eqnarray}}
\def\eqfa{\end{eqnarray}}
\def\rp#1#2{{#1\over#2}}
\def\lb#1{\label{#1}}

\def\bds#1{\boldsymbol{#1}}

\documentclass[usenatbib]{mn2e}
\usepackage{w-greek}
\usepackage{url}
\usepackage{amsmath,amscd,amssymb}
\usepackage{graphicx,epsfig}

\RequirePackage{color}

\title[Anomalous changes in the lunar orbit]{On the anomalous secular increase of the eccentricity of the orbit of the Moon}

\author[L. Iorio]{
L. Iorio$^{1}$\thanks{E-mail:
lorenzo.iorio@libero.it}\\
$^{1}$Ministero dell'Istruzione, dell'Universit\textit{\`{a}} e della Ricerca (MIUR), Viale Unit\textit{\`{a}} di Italia 68, 70125, Bari (BA), Italy
}

\begin{document}

\date{Accepted 2011 March 22. Received 2011 March 22; in original form 2011 February 1}

\maketitle

\label{firstpage}

\begin{abstract}
A recent analysis of a Lunar Laser Ranging (LLR) data record spanning 38.7 yr revealed an anomalous increase of the eccentricity $e$ of the lunar orbit amounting to $\dot e_{\rm meas}=(9\pm 3)\times 10^{-12}$ yr$^{-1}$. The present-day models of the dissipative phenomena occurring in the interiors of both the Earth and the Moon
 are not able to explain it. In this paper, we examine several dynamical effects, not modeled in the data analysis, in the framework of  long-range modified models of gravity and of the standard Newtonian/Einsteinian paradigm. It turns out that none of them can accommodate $\dot e_{\rm meas}$. Many of them do not even induce long-term changes in $e$; other models do, instead, yield such an effect, but the resulting magnitudes are in disagreement with $\dot e_{\rm meas}$. In particular,  the general relativistic gravitomagnetic acceleration of the Moon due to the Earth's angular momentum has the right order of magnitude, but the resulting Lense-Thirring secular effect for the eccentricity vanishes. A potentially viable Newtonian candidate would be a trans-Plutonian massive object (Planet X/Nemesis/Tyche) since it, actually, would affect $e$ with a non-vanishing long-term variation. On the other hand, the values for the physical and orbital parameters of such a hypothetical body required to obtain at least the right order of magnitude for $\dot e$ are completely unrealistic{: suffices it to say that an Earth-sized planet would be at 30 au, while a jovian mass would be at 200 au.} Thus, the issue of finding a satisfactorily explanation for the anomalous behavior of the Moon's eccentricity remains open.
\end{abstract}


\section{Introduction}
\citet{And010}, { in a  review of some astrometric anomalies recently detected in the solar system by several independent groups, mentioned also} an anomalous secular increase of the eccentricity\footnote{It is a dimensionless numerical parameter for which $0\leq e < 1$ holds. It  determines the shape of the Keplerian ellipse: $e=0$ corresponds to a circle, while values close to unity yield highly elongated orbits.} $e$ of the orbit of the Moon
\eqi \dot e_{\rm meas} = (9\pm 3)\times 10^{-12}\ {\rm yr^{-1}}\lb{ecce_meas}\eqf based on an analysis of a long LLR data record spanning 38.7 yr (16 March 1970-22 November 2008) performed by \citet{WiBo} with the  suite of accurate dynamical force models of the DE421 ephemerides \citep{de421a,de421b} including all relevant Newtonian and Einsteinian effects. Notice that \rfr{ecce_meas} is statistically significant at a $3\sigma-$level.
{The first presentation of such an effect appeared in \citet{JGR}, in which an extensive discussion of the state-of-the-art in modeling the tidal dissipation in both the Earth and the Moon was given. Later, \citet{hjk}, relying upon \citet{JGR}, yielded an anomalous eccentricity rate as large as $\dot e_{\rm meas}=(1.6\pm 0.5)\times 10^{-11}$ yr$^{-1}$. }
\citet{And010} {commented} that \rfr{ecce_meas} is not compatible with present, standard knowledge of dissipative processes in
the interiors of both the Earth and Moon, which were, actually, modeled by \citet{WiBo}.  The relevant physical and {osculating} orbital parameters of the Earth and the Moon are reported in Table \ref{tavola}.
\begin{table*}
\caption{Relevant physical and {osculating} orbital parameters of the Earth-Moon system. $a$ is the semimajor axis. $e$ is the eccentricity. The inclination $I$ {is referred} to the mean ecliptic at J2000.0. $\Omega$ is the longitude of the ascending node and is referred to the mean equinox and ecliptic at J2000.0. $\omega$ is the argument of pericenter. $G$ is the Newtonian gravitational constant.  The masses of the Earth and the Moon are $M$ and $m$, respectively. The orbital parameters of the Moon were retrieved from the WEB interface HORIZONS (Author: J. Giorgini. Site Manager: D. K. Yeomans.
Webmaster: A. B. Chamberlin), by JPL, NASA, at  the epoch J2000.0.
}\label{tavola}
\begin{tabular}{lllllll}
\hline
$a$ (m) & $e$ & $I$ (deg) & $\Omega$ (deg) & $\omega$ (deg) & $GM$ (m$^3$ s$^{-2}$) & $m/M$  \\
\hline
$3.81219\times 10^8$ & $0.0647$ & $5.24$ & $123.98$ & $-51.86$ & $3.98600\times 10^{14}$ &  $0.012$ \\
\hline
\end{tabular}
\end{table*}

In this paper  we look for a possible candidate for explaining such an anomaly in terms of both Newtonian and non-Newtonian gravitational dynamical effects, general relativistic or not.

To this aim, let us make the following, preliminary remarks.
Naive, dimensional evaluations of the effect caused on $e$ by an additional anomalous acceleration $A$ can be made by noticing that
\eqi \dot e\approx \rp{A}{na},\lb{misl}\eqf with
\eqi na = 1.0\times 10^3\ {\rm m\ s^{-1}}=3.2\times 10^{10}\ {\rm m\ yr^{-1}}\eqf
for the geocentric orbit of the Moon. In it, $a$ is the orbital semimajor axis, while $n\doteq\sqrt{\mu/a^3}$ is the Keplerian mean motion in which $\mu\doteq GM(1+m/M)$ is the gravitational parameter of the Earth-Moon system: $G$ is the Newtonian constant of gravitation. It turns out that an extra-acceleration  as large as  \eqi A\approx 3\times 10^{-16}\ {\rm m\ s}^{-2}=0.3\ {\rm m\ yr^{-2}}\lb{mutuu}\eqf would satisfy \rfr{ecce_meas}. In fact, a mere order-of-magnitude  analysis based on \rfr{misl} would be insufficient to draw meaningful conclusions:  finding simply that this or that dynamical effect induces an extra-acceleration of the right order of magnitude may be highly misleading. Indeed,  exact calculations of the secular variation of $e$ caused by such putative promising candidate extra-accelerations $A$ must be performed with standard perturbative techniques in order to check if they, actually, cause an averaged non-zero change of the eccentricity. Moreover, also in such potentially favorable cases caution is still in order. Indeed, it may well happen, in principle, that the resulting analytical expression for $\left\langle\dot e\right\rangle$ retains multiplicative factors\footnote{Here $e$ denotes the eccentricity: it is not the Napier number.} $1/e^k, k=1,2,3,...$ or $e^k, k=1,2,3...$ which would notably alter the size of the found non-zero secular change of the eccentricity  with respect to the expected values according to \rfr{misl}.

The plan of the paper is as follows. In Section \ref{exo} we deal with several long-range models of modified gravity. Section \ref{sta} analyzes some dynamical effects in terms of the standard Newtonian/Einsteinian laws of gravitation. The conclusions are in Section \ref{conclu}.
\section{Exotic models of modified gravity}\lb{exo}
\subsection{A Rindler-type acceleration}
As a practical example of the aforementioned caveat, let us consider the effective model for gravity of a
central object of mass $M$ at large scales recently constructed by \citet{Gru}. Among other things, it predicts the existence of a constant and uniform acceleration \eqi\bds A = A_{\rm Rin}\bds{\hat{r}}\lb{Arin}\eqf radially directed towards $M$. As shown in \citet{IorRin}, the Earth-Moon range residuals $\delta\rho$ over $\Delta t=20$ yr yield the following constrain for a terrestrial Rindler-type extra-acceleration
\eqi A_{\rm Rin}\lesssim 5\times 10^{-16}\ {\rm m\ s^{-2}}=0.5\ {\rm m\ yr^{-2}},\eqf which is in good agreement with \rfr{mutuu}.

The problem is that, actually, a radial and constant acceleration like that of \rfr{Arin} does not induce any secular variation of the eccentricity. Indeed, from the standard Gauss\footnote{It is just the case to remind that the Gauss perturbative equations are valid for any kind of perturbing acceleration $\bds A$, whatever its physical origin may be. } perturbation equation for  $e$ \citep{BeFa}
\eqi\dert e t   =  \rp{\sqrt{1-e^2}}{na}\left\{A_R\sin f + A_{T}\left[\cos f + \rp{1}{e}\left(1 - \rp{r}{a}\right)\right]\right\}, \lb{gauss}\eqf in which $f$ is the true anomaly\footnote{It is an angle counted from the pericenter, i.e. the point of closest approach to the central body, which instantaneously reckons the position of the test particle along its  Keplerian ellipse. }, and $A_R,A_T$ are the radial and transverse components of the perturbing acceleration $\bds A$, it turns out \citep{IorRin}
\eqi \Delta e = -\rp{A_{\rm Rin}\left(1-e^2\right)\left(\cos E -\cos E_0\right)}{n^2},\eqf where $E$ is the eccentric anomaly\footnote{Basically, $E$ can be regarded as a parametrization
of the polar angle in the orbital plane.}, so that
\eqi\left.\Delta e\right|_{0}^{2\pi}=0.\eqf
 \subsection{A Yukawa-type long-range modification of gravity}
 It is well known that a variety of theoretical paradigms \citep{Adel,Berto} allow for Yukawa-like deviations  from the usual Newtonian inverse-square law of gravitation \citep{Bur}. The Yukawa-type correction to the  Newtonian gravitational potential $U_{\rm N}=-\mu/r$, where $\mu\doteq GM$ is the gravitational parameter of the central body  which acts as source of the supposedly modified gravitational field,
 is \eqi  U_{\rm Y}=-\rp{\alpha\mu_{\infty}}{r}\exp\left(-\rp{r}{\uplambda}\right),\lb{uiu}\eqf where $\mu_{\infty}$ is the gravitational parameter evaluated at distances $r$ much larger than the scale length $\uplambda$.

 In order to compute the long-term effects of \rfr{uiu} on the eccentricity of a test particle it is convenient to adopt the Lagrange perturbative scheme \citep{BeFa}.
 In such a framework,  the equation for the long-term variation of $e$ is \citep{BeFa}
 \eqi \left\langle\rp{d e}{dt}\right\rangle = \rp{1}{na^2}\left(\rp{1-e^2}{e}\right)\left( \rp{1}{\sqrt{1-e^2}}\rp{\partial \mathcal{R}}{\partial\omega} -  \rp{\partial \mathcal{R}}{\partial\mathcal{M}} \right),\lb{eLag}\eqf where $\omega$ is the argument of pericenter\footnote{It is an angle in the orbital plane reckoning the position of the point of closest approach with respect to the line of the nodes which is the intersection of the orbital plane with the  reference $\left\{x,y\right\}$ plane.},  $\mathcal{M}\doteq n(t-t_p)=E -e \sin E$ is the mean anomaly of the test particle\footnote{$t_p$ is the time of passage at pericenter.}, and $\mathcal{R}$ denotes the average of the perturbing potential over one orbital revolution.
 In the case of a Yukawa-type perturbation\footnote{{Several investigations of Yukawa-type effects on the lunar data, yielding more and more tight constraints on its parameters, are present in the literature: see, e.g., \citet{Mull,lunayuk0,lunayuk1,lunayuk2}.}}, \rfr{uiu} yields
 \eqi\left\langle  U_{\rm Y}\right\rangle=-\rp{\alpha \mu_{\infty}\exp\left(-\rp{a}{\uplambda}\right)}{a}I_0\left(\rp{ae}{\uplambda}\right), \lb{poti}\eqf
where $I_0(x)$ is the modified Bessel function of the first kind
%\footnote{See on the WEB http://mathworld.wolfram.com/ModifiedBesselFunctionoftheFirstKind.html and references therein.}
$I_k(x)$ for $k=0$. An inspection of \rfr{eLag} and \rfr{poti}  immediately tells us that there is no secular variation of $e$ caused by an anomalous Yukawa-type perturbation which, thus, cannot explain \rfr{ecce_meas}.
\subsection{Other long-range exotic models of gravity}
The previous analysis has the merit of elucidating certain general features pertaining {to} a vast category of long-range modified models of gravity. Indeed, \rfr{eLag} tells us that a long-term change of $e$ occurs only if the averaged extra-potential considered explicitly depends on $\omega$ and on time through $\mathcal{M}$ or, equivalently, $E$. Actually, the  anomalous potentials  arising in the majority of long-range modified models of gravity are time-independent and spherically symmetric \citep{DGP1,Capoz1,Capoz2,DGP2,Kerr,Alle,Gruz,Jaek1,Jaek2,Nava1,Rey,Apo,Bro,Capoz3,Jaek3,Jaek4,Moff,Nava2,Nava3,San,Adk1,Adk2,Berto07,Capoz4,Capoz5,Noji,Berto09,deFe,Rug,Sot,Fab}. Anomalous accelerations $\bds A$ exhibiting a dependence on the test particle's velocity $\bds v$ were also proposed in different frameworks \citep{Jaek1,Jaek2,Hora1,Hora2,KS}. Since they have to be evaluated onto the unperturbed Keplerian ellipse, for which the following relations hold \citep{Murr010}
\begin{equation}
\left\{
\begin{array}{lll}
r & = &a\left(1-e\cos E\right), \\ \\
dt & = & \left(\rp{1-e\cos E}{n}\right)dE, \\ \\
v_R & = & \rp{nae\sin E}{1-e\cos E}, \\ \\
v_T & = & \rp{na\sqrt{1-e^2}}{1-e\cos E},
\end{array}
\right.
\end{equation}
where $v_R$ and $v_T$ are the unperturbed, Keplerian radial and transverse components of $\bds v$,
 it was straightforward to infer from  \rfr{gauss}
  that no long-term variations of the eccentricity arose at all \citep{Fop,Rugi}.

An example of time-dependent anomalous potentials occurs if either a secular change of the Newtonian gravitational constant\footnote{According to \citet{PAM},  $G$ should decrease with the age of the Universe.} \citep{Milne,PAM} or of the mass of the central body is postulated, so that a percent time variation $\dot\mu/\mu$ of the gravitational parameter can be considered. In such a case, it was recently shown with the Gauss perturbative scheme that the eccentricity experiences a secular change given by \citep{IorSyr}
\eqi \left\langle\dot e \right\rangle = \left(1+e\right)\left(\rp{\dot\mu}{\mu}\right).\lb{dote}\eqf
As remarked in \citet{IorSyr}, \rfr{ecce_meas} and \rfr{dote} would imply an increase
\eqi \rp{\dot\mu}{\mu}= +8.5\times 10^{-12}\ {\rm yr^{-1}}.\lb{MUD}\eqf If attributed to a change in $G$, \rfr{MUD} would be one order of magnitude
larger than the present-day bounds on $\dot G/G$ obtained from\footnote{{Because of the secular tidal effects, the LLR-based determinations of $\dot G$ depend more strongly on the solar perturbations, and the $\dot\mu/\mu$ values should be interpreted as being sensitive to changes in the Sun's gravitational parameter $GM$.}} LLR
\citep{Mull,Willi}. Moreover, \citet{Pit}   recently obtained a secular decrease of $G$  as large as
\eqi \rp{\dot G}{G}=(-5.9\pm 4.4)\times 10^{-14}\ {\rm yr^{-1}}\eqf
from planetary data analyses: if applied to \rfr{dote}, it is clearly insufficient to explain the empirical result of \rfr{ecce_meas}.
Putting aside a variation of $G$, the gravitational parameter of the Earth may experience a time variation because of a steady mass accretion of non-annihilating Dark Matter \citep{Blin,Khlo1,Khlo2,Foot,Adler}. \citet{Khri} and \citet{Xu} assume for the Earth
\eqi \rp{\dot M}{M}\approx +10^{-17}\ {\rm yr^{-1}},\eqf which is far smaller than \rfr{MUD}, as noticed by \citet{Iormirror}. \citet{Adler} yields an even smaller figure for $\dot M/M$.
\section{Standard Newtonian and Einsteinian dynamical effects}\lb{sta}
In this Section we look at possible dynamical causes for \rfr{ecce_meas} in terms of standard Newtonian and general relativistic gravitational effects which were not modeled in processing the LLR data.
\subsection{The general relativistic Lense-Thirring field and other stationary spin-dependent effects}
It is interesting to notice that the magnitude of the general relativistic \citet{LT} acceleration experienced by the Moon because of the Earth's angular momentum $S=5.86\times 10^{33}$ kg m$^2$ s$^{-1}$ \citep{IERS} is
just
\eqi A_{\rm LT}\approx \rp{2v G S}{c^2 a^3}=1.6\times 10^{-16}\ {\rm m\ s^{-2}} = 0.16\ {\rm m\ yr^{-2}},\eqf i.e. close to \rfr{mutuu}.
On the other hand, it is well known that the  Lense-Thirring effect does not cause long-term variations of the eccentricity. Indeed, {the integrated shift of $e$ from an initial epoch corresponding to $f_0$ to a generic time corresponding to $f$ is } \citep{Soff}
\eqi \Delta e = -\rp{2 G S\cos I{^{'}}\left(\cos f-\cos f_0\right)}{c^2 n a^3\sqrt{1-e^2}}.\lb{letie}\eqf {From \rfr{letie} it straightforwardly follows that after one orbital revolution, i.e. for $f\rightarrow f_0+2\pi$, the gravitomagnetic shift of $e$ vanishes.}
In fact, \rfr{letie} holds only for a specific orientation of $\bds S$, which is assumed to be directed along the reference $z$ axis{; incidentally, let us remark that, in this case, the angle $I^{'}$ in \rfr{letie} is to be intended as the inclination of the Moon's orbit with respect to the Earth's equator\footnote{{It approximately varies between 18 deg and 29 deg \citep{seidel,hjk}}.}}  Actually, in \citet{Iorfar} it was shown that $e$ does not secularly change also for a generic orientation of $\bds S$ since\footnote{{Here $I,\Omega$ can be thought as referring to the mean ecliptic at J2000.0. Generally speaking, the longitude of the ascending node $\Omega$ is an angle in the reference $\{x,y\}$ plane determining the position of  the line of the nodes with respect to the reference $x$ direction.}}
\eqi \mathcal{R}_{\rm LT}=\rp{2G n}{c^2 a (1-e^2)}\left[S_z\cos I+\sin I\left(S_x\sin\Omega-S_y\cos\Omega\right)\right].\eqf
Thus, standard general relativistic gravitomagnetism cannot be the cause of \rfr{ecce_meas}.

\citet{Iorkds} explicitly worked out the gravitomagnetic orbital effects induced on the trajectory of a test
particle by the the weak-field approximation of the Kerr-de Sitter metric. No long-term variations for $e$ occur.
Also the general relativistic spin-spin effects \textit{\`{a}} \textit{la} Stern-Gerlach  do not cause long-term variations in the eccentricity \citep{Iorfar}.
\subsection{General relativistic gravitomagnetic time-varying effects}
By using the Gauss perturbative equations, \citet{RugIor}  analytically worked out the long-term variations of all the Keplerian orbital elements caused by general relativistic gravitomagnetic time-varying effects.
For the eccentricity, \citet{RugIor} found a non-vanishing secular change given by
\eqi \left\langle\dot e\right\rangle = -\rp{G S_1\left(2+e\right)\cos I{^{'}}}{c^2 a^3 n},\lb{spind}\eqf in which
$S_1$ denotes a linear change of the magnitude of the angular momentum of the central rotating body.

In the case of the Earth, \citet{RugIor} quote
\eqi S_1 = -5.6\times 10^{16}\ {\rm kg\ m^2\ s^{-2}}\lb{spinf}\eqf
due to the secular decrease of the Earth's diurnal rotation period \citep{Bros} $\dot P/P = -3\times 10^{-10}$ yr$^{-1}$.
Thus, \rfr{spind} and \rfr{spinf} yield for the Moon's eccentricity
\eqi \left\langle\dot e\right\rangle =  -2\times 10^{-23}\ {\rm yr^{-1}},\eqf
which is totally negligible with respect to \rfr{ecce_meas}.
\subsection{The first and second post-Newtonian static components of the gravitational field}
Also the first post-Newtonian,  Schwarzschild-type, spherically symmetric static component of the gravitational field, which was, in fact, fully modeled by \citet{WiBo}, does not induce long-term variations of $e$ \citep{Soff}. The same holds also for the spherically  symmetric second post-Newtonian terms of order $\mathcal{O}(c^{-4})$ \citep{2pn1,2pn2,2pn3}, which were not modeled by \citet{WiBo}. Indeed, let us recall that the components of the spacetime metric tensor $g_{\mu\nu},\ \mu,\nu=0,1,2,3$, are, up to the second post-Newtonian order, \citep{Nordt}
\begin{equation}
\left\{
\begin{array}{lll}
g_{00}& \cong & 1-2\rp{\mathfrak{M}}{r}+2\left(\rp{\mathfrak{M}}{r}\right)^2 -\rp{3}{2}\left(\rp{\mathfrak{M}}{r}\right)^3 +\ldots, \\ \\
g_{ij} & \cong &-\delta_{ij}\left[1+2\rp{\mathfrak{M}}{r} +\rp{3}{2}\left(\rp{\mathfrak{M}}{r}\right)^2 + \ldots\right], \ i,j=1,2,3, \lb{2PN}
\end{array}
\right.
\end{equation}
where $\mathfrak{M}\doteq\mu/c^2$. Notice that \rfr{2PN} are written in the standard isotropic gauge, suitable for a direct comparison with the observations.
Incidentally, let us remark that the second post-Newtonian acceleration for the Moon is just
\eqi A_{\rm 2PN} \approx \rp{\mu^2 n^2}{c^4 r}=4\times 10^{-25}\ {\rm m\ s^{-2}}=4\times 10^{-10}\ {\rm m\ yr^{-2}}.\eqf
\subsection{The general relativistic effects for an oblate body}
\citet{Sof1}, by using the Gauss perturbative scheme and the usual Keplerian orbital elements,  analytically worked out  the first-order post-Newtonian orbital effects in the field of an oblate body with adimensional quadrupole mass moment $J_2$ and equatorial radius $R$.

It turns out that the eccentricity undergoes a non-vanishing {harmonic} long-term variation which, in general relativity, is\footnote{{Here $\omega^{'}$ refers to the Earth's equator, so that its period amounts to $8.85$ yr \citep{luna}.}} \citep{Sof1}
\eqi\left\langle\dot e\right\rangle = \rp{21 n J_2 e \sin^2 I{^{'}}}{8\left(1-e^2\right)^3}\left(\rp{R}{a}\right)^2\left(\rp{\mu}{c^2 a}\right)\left(1+\rp{e^2}{2}\right)\sin2\omega{^{'}}.\lb{j2c}\eqf

In view of the fact that, for the Earth, it is $J_2=1.08263\times 10^{-3}$ \citep{IERS} and $R=6.378\times 10^3$ m \citep{IERS}, it turns out that
the first-order general relativistic $J_2 c^{-2}$ effect is not capable to explain \rfr{ecce_meas} since {it is}
\eqi \left\langle\dot e\right\rangle{\lesssim} 4\times 10^{-19}\ {\rm yr^{-1}}\eqf {as a limiting value for the periodic perturbation of \rfr{j2c}.}

\citet{Sof1} pointed out that the second-order mixed perturbations due to the Newtonian quadrupole field and the general relativistic Schwarzschild acceleration are of the same order of magnitude of the first-order ones: their orbital effects were analytically worked out by  \citet{Heim} with the technique of the canonical Lie transformations applied to the Delaunay variables. Given their negligible magnitude, we do not further deal with them.
\subsection{A massive ring of minor bodies}
A Newtonian effect which was not modeled is the action of  the Trans-Neptunian Objects (TNOs) of the Edgeworth-Kuiper belt \citep{Edge,Kui}. It can be taken into account by means of a massive circular ring having mass $m_{\rm ring}\leq 5.26\times 10^{-8}\ {\rm M}_{\odot}$ \citep{Pit} and radius $R_{\rm ring}=43$ au \citep{Pit}.
Following \citet{Fie},
it causes a perturbing radial acceleration
\eqi \bds{A}_{\rm ring}=\rp{Gm_{\rm ring}}{2 r R^2_{\rm ring}}\left[b^{(1)}_{\rp{3}{2}}(\alpha) -\alpha b^{(0)}_{\rp{3}{2}}(\alpha)\right]\bds r,\ \alpha\doteq \rp{r}{R_{\rm ring}}.\lb{aringa}\eqf
The Laplace coefficients are defined as \citep{Murr}
\eqi b_s^{(j)}\doteq \rp{1}{\pi}\int_0^{2\pi}\rp{\cos(j\psi)}{\left(1-2\alpha\cos\psi +\alpha^2\right)^s}d\psi,\eqf
where $s$ is a half-integer. Since for the Moon $\alpha \approx 3\times 10^{-10}$, \rfr{aringa} becomes
\eqi \bds{A}_{\rm ring}\approx \rp{Gm_{\rm ring}}{2 r R^2_{\rm ring}}\alpha\bds r,\eqf
with
\eqi A_{\rm ring} \approx 10^{-23}\ {\rm m\ s^{-2}} \approx 10^{-8}\ {\rm m\ yr^{-{2}}}, \lb{frunzo}\eqf which is far smaller than \rfr{mutuu}.
% Anyway, it can be shown by using the Gauss perturbative equation of \rfr{gauss} that the long-term variation of $e$ caused by \rfr{aringa} vanishes.

{Actually, the previous results holds, strictly speaking, in a heliocentric frame since the distribution of the TNOs is assumed to be circular with respect to the Sun. Thus, it may be argued that a rigorous geocentric calculation should take into account for the non-exact circularity of the TNOs belt with respect to the Earth. Anyway, in view of the distances involved, such departures from azimuthal symmetry would plausibly display as small corrections to the main term of \rfr{aringa}. Given the negligible orders of magnitude involved by \rfr{frunzo}, we feel it is unnecessary to perform such further calculations. }

The dynamical action of the belt of the minor asteroids \citep{Kras} was, actually, modeled, so that we do not consider it here.
\subsection{A distant massive object: Planet X/Nemesis/Tyche}
 A promising candidate for explaining the anomalous increase of the lunar eccentricity
   may be, at least in principle, a  trans-Plutonian  massive body of planetary size located in the remote peripheries of the solar system: Planet X/Nemesis/Tyche \citep{Lyk,Mel,Fer011,Mat}. Indeed, as we will see, the perturbation induced by it would actually cause a non-vanishing long-term variation of $e$. Moreover, since it depends on the spatial position of X in the sky and on its tidal parameter
\eqi \mathcal{K}_{\rm X}\doteq \rp{Gm_{\rm X}}{d_{\rm X}^3},\eqf where $m_{\rm X}$ and $d_{\rm X}$ are the mass and the distance of X, respectively, it may happen that a suitable combination of them is able to reproduce the empirical result of \rfr{ecce_meas}.

Let us recall that{, in general,} the perturbing potential {felt by a test particle orbiting a central body} due to a very distant, pointlike mass can be cast into the following quadrupolar form \citep{Hogg}
\eqi U_{\rm X} = \rp{\mathcal{K}_{\rm X}}{2}\left[r^2 -3\left(\bds r\bds\cdot\bds{\hat{l}} \right)^2\right],\lb{ux}\eqf
where
 $\bds{\hat{l}}=\left\{l_x,l_y,l_z\right\}$ is a unit vector directed towards X determining its position in the sky; its components are not  independent since the constraint
\eqi l_x^2+l_y^2+l_z^2=1\eqf holds.
By introducing the ecliptic latitude $\beta_{\rm X}$ and longitude $\lambda_{\rm X}$ {in a geocentric ecliptic frame}, it is possible to write
\begin{equation}
\left\{
\begin{array}{lll}
l_x & = & \cos\beta_{\rm X}\cos\lambda_{\rm X}, \\ \\
l_y & = & \cos\beta_{\rm X}\sin\lambda_{\rm X}, \\ \\
l_z & = & \sin\beta_{\rm X}.
\end{array}
\right.
\end{equation}
In \rfr{ux} $\bds r=\left\{x,y,z\right\}$ {is the geocentric position vector of} the perturbed particle{, which, in the present case, is the Moon}.
\citet{IorX} has recently shown that the average of \rfr{ux} over one orbital revolution {of the particle} is
\begin{equation}
\left\langle U_{\rm X} \right\rangle   =  \rp{{\mathcal{K}_{\rm X}}a^2}{32}\mathcal{U}\left(e,I,\Omega,\omega; \bds{\hat{l}}\right),\lb{us}
\end{equation}
with $\mathcal{U}\left(e,I,\Omega,\omega; \bds{\hat{l}}\right)$ given by \rfr{uavera}.
\begin{figure*}
\begin{equation}
\begin{array}{lll}
\mathcal{U} &\doteq & - \left(2 + 3 e^2\right)  \left( -8  + 9 l_x^2 + 9 l_y^2 +
        6 l_z^2\right)   -120 e^2 \sin 2\omega \left(l_x \cos\Omega + l_y \sin\Omega\right)  \left[l_z \sin I +\right. \\ \\
    &+&\left.\cos I \left(l_y \cos\Omega - l_x \sin\Omega\right) \right]  - 15 e^2 \cos 2\omega \left[3
 \left(l_x^2 - l_y^2\right)  \cos 2\Omega + 2 \left(
              l_x^2 + l_y^2 - 2
                    l_z^2\right)  \sin^2 I -\right. \\ \\
                     &-&\left. 4 l_z \sin 2I \left(l_y
\cos\Omega - l_x \sin\Omega\right)  + 6 l_x l_y \sin 2\Omega\right]  -
          6 \left(2 + 3 e^2\right)  \left[ \left(l_x^2 - l_y^2\right)   \cos 2\Omega \sin^2 I +\right. \\ \\
           &+&\left. 2 l_z \sin 2I \left(l_y
                    \cos\Omega - l_x \sin\Omega\right)  +
                2 l_x l_y \sin^2 I \sin 2\Omega\right]  - 3\cos 2I \left\{ \left(2 + 3
                    e^2\right)  \left(l_x^2 + l_y^2 - 2 l_z^2\right)  +\right. \\ \\
                     &+&\left. 5 e^2 \cos 2\omega \left[ \left(l_x^2 - l_y^2\right)   \cos 2\Omega + 2 l_x l_y \sin 2\Omega\right] \right\}.
\end{array}\lb{uavera}
\end{equation}
\end{figure*}
Note that \rfr{us} and \rfr{uavera} are exact: no approximations in $e$ were used.
In the integration  $\bds{\hat{l}}$ was kept fixed over one orbital revolution of the {Moon}, as it is reasonable given the assumed large distance of X with respect to it.

The Lagrange planetary equation  of \rfr{eLag} straightforwardly yields \citep{IorX}
\eqi\left\langle\dot e\right\rangle  = \rp{15\mathcal{K}_{\rm X}e\sqrt{1-e^2}}{16n}\mathcal{E}\left(I,\Omega,\omega; \bds{\hat{l}}\right),\lb{eccecazzo}\eqf
with $\mathcal{E}\left(I,\Omega,\omega; \bds{\hat{l}}\right)$ given by \rfr{zumba}.
\begin{figure*}
 \begin{equation}
\begin{array}{lll}
\mathcal{E} & \doteq & -8  l_z \cos 2\omega \sin I \left(l_x \cos\Omega + l_y \sin\Omega\right)  +
    4 \cos I \cos 2 \omega \left[ -2  l_x l_y \cos 2\Omega +\right.         \\ \\
    &+&\left.  \left(l_x^2 - l_y^2\right)  \sin 2\Omega\right]  +
    \sin 2\omega \left[ \left(l_x^2 - l_y^2\right)   \left(3 +
                \cos 2I\right)  \cos 2\Omega +
          2 \left(l_x^2 + l_y^2 - 2 l_z^2\right)  \sin^2 I -\right. \\ \\
          &-&\left.
          4 l_z \sin 2I \left(l_y \cos\Omega -
                l_x \sin\Omega\right)  +
          2 l_x l_y \left(3 + \cos 2I\right)  \sin 2\Omega\right].\lb{zumba}
          \end{array}
\end{equation}
\end{figure*}

Actually, the expectations concerning X are doomed to fade away. Indeed, apart from the modulation introduced by the presence of the time-varying $I,\omega$ and $\Omega$ in \rfr{zumba}, the values for the tidal parameter which would allow to obtain \rfr{ecce_meas} are too large for all the conceivable positions $\left\{\beta_{\rm X},\lambda_{\rm X}\right\}$ of X in the sky. This can easily be checked by keeping $\omega$ and $\Omega$ fixed at their J2000.0 values as a first approximation.

Figure \ref{figura} depicts the X-induced variation of the lunar eccentricity, normalized to \rfr{ecce_meas}, as a function of $\beta_{\rm X}$ and $\lambda_{\rm X}$ for the scenarios by \citet{Lyk} ($m_{\rm X}^{\rm max}=0.7\ m_{\oplus}, d_{\rm X}^{\rm min}=101.3$ au), and by \citet{Mat} ($m^{\rm max}_{\rm X}=4\ m_{\rm Jup}, d_{\rm X}=30$ kau).
\begin{figure*}
\centering
\begin{tabular}{cc}
\epsfig{file=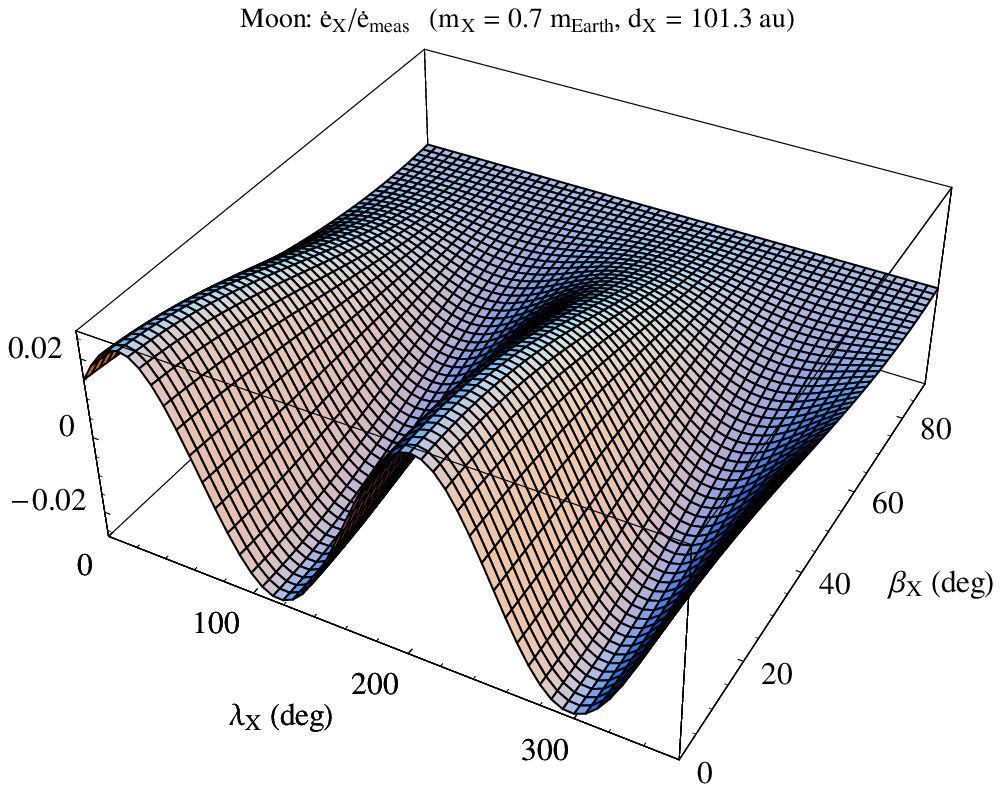,width=0.50\linewidth,clip=} & \epsfig{file=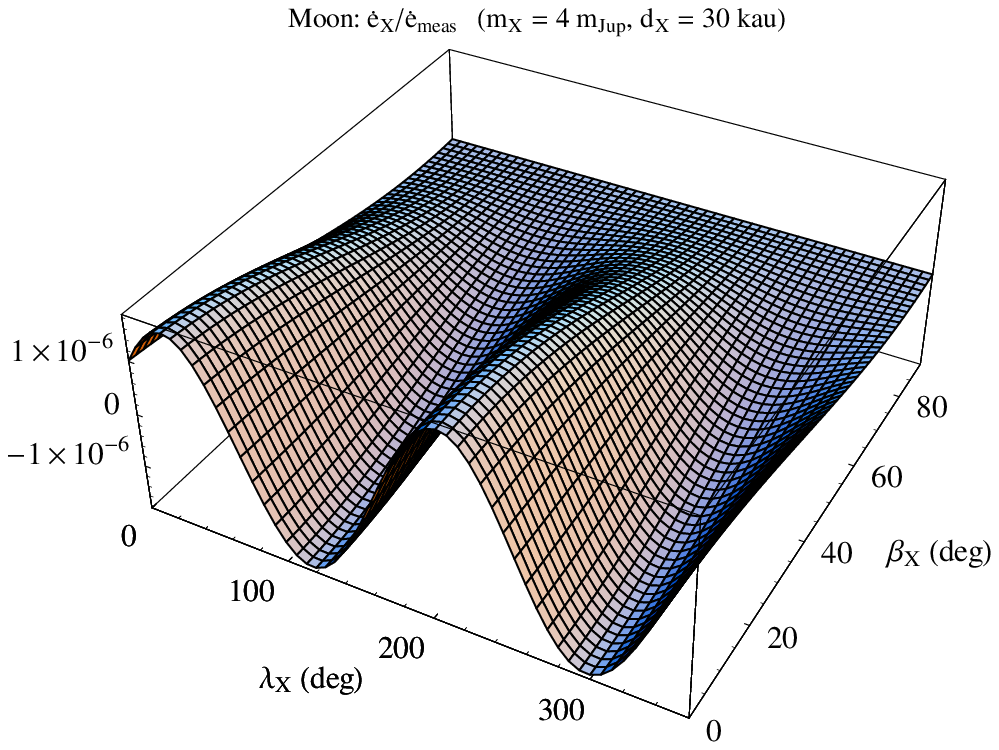,width=0.50\linewidth,clip=}
\end{tabular}
\caption{Long-term variation of the lunar eccentricity, normalized to \rfr{ecce_meas}, induced by a trans-Plutonian, pointlike object X as a function of its ecliptic latitude $\beta_{\rm X}$ and longitude $\lambda_{\rm X}$. The node $\Omega$ and the perigee $\omega$ of the Moon were kept fixed to the J2000.0 values quoted in Table \ref{tavola}. The scenarios for the perturbing body X are those by \citet{Lyk} (left panel), and by \citet{Mat} (right panel). }\lb{figura}
\end{figure*}
It can be noticed that the  physical and orbital features of X postulated by such two recent theoretical models would induce long-term variations of the lunar eccentricity much smaller than \rfr{ecce_meas}.
Conversely, it turns out that a tidal parameter as large as
\eqi \mathcal{K}_{\rm X}=4.46\times 10^{-24}\ {\rm s^{-2}}\lb{cazzara}\eqf would yield the result of \rfr{ecce_meas}. Actually, \rfr{cazzara} is totally unacceptable since it corresponds to distances of X as absurdly small as $d_{\rm X}=30$ au for a terrestrial body, and $d_{\rm X}=200$ au for a Jovian mass \citep{IorX}.

We must conclude that not even the hypothesis of Planet X is  a viable one to explain the anomalous increase of the lunar eccentricity of \rfr{ecce_meas}.
\section{Summary and conclusions}\lb{conclu}
In this paper we dealt with the anomalous increase of the eccentricity $e$ of the orbit of the Moon recently reported from an analysis of a multidecadal record of LLR data points.

We looked for possible explanations in terms of unmodeled dynamical features of motion within either the standard Newtonian/Einsteinian paradigm or several long-range models of modified gravity. As a general rule, we, first, noticed that it would be misleading to simply find the right order of magnitude for the extra-acceleration due to this or that candidate effect. Indeed, it is mandatory to explicitly check if a potentially viable candidate does actually induce a non-vanishing averaged variation of the eccentricity.
This holds, in principle, for the search of an explanation of any other possible anomalous effect.
Quite generally, it turned out that any time-independent and spherically symmetric perturbation does not  affect the eccentricity with long-term changes.

Thus, most of the  long-range modified models of gravity proposed in more or less recent times for other scopes are automatically ruled out.
%A Pioneer-like anomalous acceleration directed towards the Sun induces a long-term variation of $e$, but its magnitude is too large by several orders of %magnitude. Conversely,
The present-day limits on the magnitude of a terrestrial Rindler-type perturbing acceleration are of the right order of magnitude, but it does not secularly affect $e$. As  time-dependent candidates capable to cause secular shifts of $e$, we considered the possible variation of the  Earth's gravitational parameter $\mu$ both because of a temporal variation of the Newtonian constant of gravitation $G$ and of its mass itself due to a steady mass accretion of non-annihilating Dark Matter. In both cases, the resulting time variations of $e$ are too small by several orders of magnitude.

Moving to standard general relativity, we found that the gravitomagnetic Lense-Thirring lunar acceleration due to the Earth's angular momentum, not modeled in the data analysis, has the right order of magnitude, but it, actually, does not induce secular variations of $e$. The same holds also for other general relativistic spin-dependent effects. Conversely, $e$ undergoes long-term changes caused by the general relativistic first-order effects due to the Earth's oblateness, but they are far too small.
The second-order post-Newtonian part of the gravitational field does not affect the eccentricity.

Within the Newtonian framework, we considered the action of an almost circular massive ring modeling the Edgeworth-Kuiper belt of Trans-Neptunian Objects, but it does not induce secular variations of $e$. In principle, a viable candidate would be a putative trans-Plutonian massive object (PlanetX/Nemesis/Tyche), recently revamped to accommodate certain features of the architecture of the Kuiper belt and of the distribution of the comets in the Oort cloud, since it would cause a non-vanishing long-term variation of the eccentricity. Actually, the values for its mass and distance needed to explain the empirically determined increase of the lunar eccentricity would be highly unrealistic and in contrast with the most recent viable theoretical scenarios for the existence of such a body. For example, a terrestrial-sized body should be located at just 30 au, while an object with the mass of Jupiter should be at 200 au.

Thus, in conclusion, the issue of finding a satisfactorily explanation of the observed orbital anomaly of the Moon still remains  open.
Our analysis should have effectively restricted the field of possible explanations, indirectly pointing towards either non-gravitational, mundane effects or some artifacts in the data processing. Further data analyses, hopefully performed by independent teams, should help in shedding further light on such an astrometric anomaly.
\section*{Acknowledgements}
 I thank an anonymous referee for her/his valuable critical remarks which greatly contributed to improve the manuscript.

%-----------------------------------------

%----------


\begin{thebibliography}{}

\bibitem[\protect\citeauthoryear{Adelberger et al.}{2003}]{Adel}
Adelberger E.G.,  Heckel B.R.,  Nelson A.E., 2003, Ann. Rev. Nucl. Part. Sci., 53, 77

\bb{Adkins \& McDonnell}{2007}{Adk1}
Adkins G. S.,  McDonnell J., 2007, Phys. Rev. D., 75, 082001

\bb{Adkins et al.}{2007}{Adk2}
Adkins G. S.,  McDonnell J., Fell R. N., 2007, Phys. Rev. D., 75, 064011

\bb{Adler}{2008}{Adler}
Adler S. L., 2008, J. of Physics A: Mathematical and Theoretical, 41,  412002

\bb{Allemandi et al.}{2005}{Alle}
Allemandi G.,  Francaviglia M.,  Ruggiero M. L.,  Tartaglia A., 2005,  Gen. Relativ.  Gravit., 37, 1891


%\bb{Anderson et al.}{1998}{Pio1}
%Anderson J. D., Laing P. A., Lau E. L., Liu A. S., Nieto M. M., Turyshev
%S. G., 1998, Phys. Rev. Lett., 81, 2858

%\bb{Anderson et al.}{2002}{Pio2}
%Anderson J. D., Laing P. A., Lau E. L., Liu A. S., Nieto M. M., Turyshev
%S. G., 2002, Phys. Rev. D, 65, 082004

\bb{Anderson \& Nieto}{2010}{And010}
Anderson J. D., Nieto M. M., 2010, Astrometric solar-system anomalies. In:  Klioner S.A.,   Seidelmann P.K.,  Soffel M.H.,  (eds.)  Relativity in Fundamental Astronomy: Dynamics,
Reference Frames, and Data Analysis,
Proceedings IAU Symposium No. 261, Cambridge University Press, Cambridge, 2010, pp. 189-197

\bb{Apostolopoulos \& Tetradis}{2006}{Apo}
Apostolopoulos P. S.,  Tetradis N., 2006,
Phys. Rev. D, 74, 064021

\bibitem[\protect\citeauthoryear{Bertolami \& P\'{a}ramos}{2005}]{Berto}
Bertolami O., P\'{a}ramos J., 2005, Phys. Rev. D, 71, 023521


\bb{Bertolami et al.}{2007}{Berto07}	
Bertolami O., B\"{o}hmer C. G., Harko T., Lobo F. S. N., 2007, Phys. Rev. D, 75,  104016
	

\bb{Bertolami \& Santos}{2009}{Berto09}
Bertolami O., Santos N. M. C., 2009, Phys. Rev. D, 79, 127702

\bb{Bertotti et al.}{2003}{BeFa}
Bertotti B., Farinella P., Vokrouhlick\'{y} D., 2003, Physics of the Solar System, Kluwer Academic Press, Dordrecht

\bb{Blinnikov \& Khlopov}{1983}{Blin}
Blinnikov S. I., Khlopov M. Yu., 1983,  Sov.
Astron., 27, 371.

\bb{Brosche \& Schuh}{1998}{Bros}
Brosche P., Schuh H., 1998, Surv. Geophys., 19, 417

\bb{Brownstein \&  Moffat}{2006}{Bro}
Brownstein J. R.,  Moffat J. W., 2006,  Class. Quantum Grav., 23, 3427


\bibitem[\protect\citeauthoryear{Burgess \& Cloutier}{1988}]{Bur}
Burgess C.P., Cloutier J., 1988, Phys. Rev. D, 38, 2944

\bb{Capozziello}{2007}{Capoz4}
Capozziello S., 2007, Int. J.
Geom. Meth. Mod. Phys.,  4,  53

\bb{Capozziello et al.}{2001}{Capoz1}
Capozziello S.,  De Martino S.,  De Siena S.,  Illuminati F., 2001, Mod. Phys. Lett. A, 16, 693

\bb{Capozziello \& Lambiase}{2003}{Capoz2}
Capozziello S.,  Lambiase G., 2003, Int. J. Mod. Phys. D, 12, 843

\bb{Capozziello et al.}{2006}{Capoz3}
Capozziello S.,  Cardone V. F.,  Francaviglia M.,  2006, Gen. Relativ.  Gravit.,  38,  711

\bb{Capozziello \& Francaviglia}{2008}{Capoz5}
Capozziello S.,  Francaviglia M., 2008, Gen. Relativ. Gravit., 40, 357

\bb{Damour \& Sch\"{a}fer}{1988}{2pn1}
Damour T.,  Sch\"{a}fer G., 1988, Nuovo Cimento B, 101, 127

\bb{de Felice \& Tsujikawa}{2010}{deFe}
de Felice A., Tsujikawa S., 2010, Living Rev. Rel., 13, 3

\bb{Dirac}{1937}{PAM}
Dirac P. A. M., 1937,  Nature, 139, 323

\bb{Dvali et al.}{2000}{DGP1}
Dvali G.,  Gabadadze G.,  Porrati M., 2000, Phys. Lett. B, 485, 208

\bb{Dvali et al.}{2003}{DGP2}
Dvali G., Gruzinov A.,  Zaldarriaga M., 2003, Phys. Rev. D, 68, 024012

\bb{Edgeworth}{1943}{Edge}
Edgeworth K. E., 1943, J. Brit. Astron. Assoc., 53, 181

\bb{Fabrina et al.}{2011}{Fab}	
Farina C., Kort-Kamp W. J. M., Mauro Filho S., Shapiro I. L., 2011, arXiv:1101.5611

\bb{Fern\'{a}ndez}{2011}{Fer011}	
Fern\'{a}ndez J.A., 2011, ApJ, 726, 33

\bb{Fienga et al.}{2008}{Fie}
Fienga A.,  Manche H.,  Laskar J.,  Gastineau M., 2008,   Astron. Astrophys., 477, 315

\bb{Folkner et al.}{2008}{de421a}
Folkner W. M.,  Williams J. G.,  Boggs D. H., 2008, The Planetary and Lunar Ephemeris DE 421, JPL IOM 343R-08-003

\bb{Foot}{2004}{Foot}
Foot R., 2004,  Int. J. Mod. Phys. D, 13, 2161

\bb{Grumiller}{2010}{Gru}
 Grumiller D., 2010, Phys. Rev.
Lett., 105, 211303

\bb{Gruzinov}{2005}{Gruz}
Gruzinov A., 2005, New Astron., 10, 311

\bb{Heimberger et al.}{1990}{Heim}
Heimberger J., Soffel M., Ruder H., 1990, Celest. Mech. Dyn. Astron., 47, 205

\bb{Hogg et al.}{1991}{Hogg}
Hogg D., Quinlan G., Tremaine S., 1991, AJ, 101, 2274


\bb{Ho\v{r}ava}{2009a}{Hora1}
Ho\v{r}ava P., 2009a, Phys. Rev. D, 79, 84008

\bb{Ho\v{r}ava}{2009b}{Hora2}
Ho\v{r}ava P., 2009b, Phys. Rev. Lett. 102, 161301



\bb{Iorio}{2007}{Fop}
Iorio L., 2007, Found. Phys., 73, 897

%\bb{Iorio}{2009}{Pio}
%Iorio L., 2009, Int. J. Mod. Phys. D., 18, 947

\bb{Iorio}{2010a}{IorRin}
Iorio L., 2010a, arXiv:1012.0226

\bb{Iorio}{2010b}{IorSyr}
Iorio L., 2010b, Schol. Res. Exch., 2010, 261249

\bb{Iorio}{2010c}{Iormirror}
Iorio L., 2010c, J. Cosmol. Astropart. Phys., 05, 018

\bb{Iorio}{2010d}{Iorfar}
Iorio L., 2010d, arXiv:1012.5622


\bb{Iorio}{2011}{IorX}
Iorio L., 2011, arXiv:1101.2634

\bb{Iorio \& Ruggiero}{2009}{Iorkds}
Iorio L., Ruggiero M. L., 2009, J. Cosmol. Astropart. Phys., 03, 024

\bb{Iorio \& Ruggiero}{2010}{Rugi}
Iorio L., Ruggiero M. L., 2010,  Int. J. Mod. Phys. A., 25, 5399

\bb{Jaekel \& Reynaud}{2005a}{Jaek1}
Jaekel M.-T.,  Reynaud S., 2005a, Mod. Phys. Lett. A, 20, 1047

\bb{Jaekel \& Reynaud}{2005b}{Jaek2}
Jaekel M.-T.,  Reynaud S., 2005b,  Class. Quantum Grav., 22, 2135



\bb{Jaekel \& Reynaud}{2006a}{Jaek3}
Jaekel M.-T.,  Reynaud S., 2006a, Class. Quantum Grav., 23, 777

\bb{Jaekel \& Reynaud}{2006b}{Jaek4}
Jaekel M.-T.,  Reynaud S., 2006b, Class. Quantum Grav., 23, 7561


\bb{Kehagias \& Sfetsos}{2009}{KS}
Kehagias A., Sfetsos K., 2009, Phys. Lett. B, 678, 123

\bb{Kerr et al.}{2003}{Kerr}
Kerr A. W., Hauck J. C.,  Mashhoon B., 2003, Class. Quantum Grav., 20, 2727

\bb{Khlopov et al.}{1991}{Khlo1}
Khlopov M. Yu.,  Beslin G. M.,  Bochkarev N. G.,  Pustil'nik L. A.,  Pustil'nik S. A., 1991,
 Sov. Astron., 35, 21

\bb{Khlopov}{1999}{Khlo2}
Khlopov M. Yu., 1999, Cosmoparticle physics, World Scientific, Singapore

\bb{Krasinsky et al.}{2002}{Kras}
Krasinky G. A., Pitjeva E. V., Vasilyev M. V., Yagudina E. I., 2002,
 Icarus, 158, 98

\bb{Khriplovich \& Shepelyansky}{2009}{Khri}
Khriplovich I. B., Shepelyansky D. L., 2009,
Int. J. Mod. Phys. D, 18, 1903

\bb{Kuiper}{1951}{Kui}
Kuiper G. P., 1951, in Hynek J. A., ed., Proc. Topical Symp. Commemorating
the 50th Anniversary of the Yerkes Observatory and Half a Century of
Progress in Astrophysics. McGraw-Hill, New York, p. 357

%\bb{L\"{a}mmerzahl}{2007}{Lamm}
%L\"{a}mmerzahl C., 2007, personal communication to E.M. Standish, quoted in \citet{Sta}, p. 261

\bibitem[\protect\citeauthoryear{Lense \& Thirring}{1918}]{LT}
Lense J.,  Thirring H., 1918,
Phys. Z.,
%} {\bf
19,
%},
156

\bb{Lykawka \& Mukai}{2008}{Lyk}
Lykawka P.S., Mukai T., 2008, AJ, 135, 1161

\bb{Matese \& Whitmire}{2011}{Mat}
Matese J.J., Whitmire D.P., 2011, Icarus, 211, 926

\bibitem[\protect\citeauthoryear{McCarthy \& Petit}{2004}]{IERS}
McCarthy D.D., Petit G., 2004,  IERS Technical Note  No. 32. IERS Conventions (2003). 12. Frankfurt am Main:  Verlag des Bundesamtes f\"{u}r Kartographie und Geod\"{a}sie

\bb{Melott \& Bambach}{2010}{Mel}
Melott A.L., Bambach R.K., 2010, MNRAS, 407,  L99

\bb{Milne}{1935}{Milne}
Milne E. A., 1935, Relativity, Gravity and World Structure. Oxford: Clarendon Press

\bb{Moffat}{2006}{Moff}
Moffat J. W., 2006, J. Cosmol. Astropart. Phys., 0603, 004

\bb{M\"{u}ller \& Biskupek}{2007}{Mull}
M\"{u}ller J.,  Biskupek L., 2007,   Class. Quantum
Gravit., 24,  4533

{
\bb{M\"{u}ller et al.}{2007}{lunayuk0}
M\"{u}ller J., Williams J. G., Turyshev S. G., Shelus P. J., 2007, Potential Capabilities of Lunar Laser Ranging for Geodesy and Relativity. In: Tregoning P., Rizos C. (eds.) Dynamic Planet. International Association of Geodesy Symposia Volume 130, Part IX, Springer, Berlin, pp. 903-909
}

{
\bb{M\"{u}ller et al.}{2008}{lunayuk1}
M\"{u}ller J., Williams J. G., Turyshev S. G., 2008, Lunar Laser Ranging Contributions to Relativity and Geodesy. In: Dittus H.,  L\"{a}mmerzahl C.,  Turyshev S. G. (eds.) Lasers, Clocks and Drag-Free Control.
Astrophysics and Space Science Library Volume 349, Springer, Berlin, pp. 457-472
}

{
\bb{M\"{u}ller et al.}{2009}{lunayuk2}
M\"{u}ller J., Biskupek L., Oberst J., Schreiber U., 2009, Contribution of Lunar Laser Ranging to Realise Geodetic Reference Systems. In: Drewes H. (ed.) Geodetic Reference Frames. International Association of Geodesy Symposia  Volume 134, Part 2, Springer, Berlin, pp. 55-59,
}


\bb{Murray \& Dermott}{1999}{Murr}
Murray C. D.,  Dermott S. F., 1999, Solar System Dynamics, Cambridge University Press,
Cambridge

\bb{Murray \& Correia}{2010}{Murr010}
Murray C.D., Correia A.C.M., 2010, Keplerian Orbits and Dynamics. In:  Seager S. (ed.), Exoplanets, University of Arizona Press, Tucson, pp. 15-23

\bb{Navarro \& van Acoleyen}{2005}{Nava1}
Navarro I.,   van Acoleyen K., 2005,
Phys. Lett. B, 622, 1

\bb{Navarro \& van Acoleyen}{2006a}{Nava2}
Navarro I.,   van Acoleyen K., 2006a, J.  Cosmol.  Astropart. Phys., 008, 003


\bb{Navarro \& van Acoleyen}{2006b}{Nava3}
Navarro I.,   van Acoleyen K., 2006b,
J. Cosmol. Astropart. Phys., 006, 009

\bb{Nojiri \& Odintsov}{2007}{Noji}
Nojiri S.,  Odintsov S. D., 2007,  Int. J.  Geom. Meth.  Mod. Phys., 4,  115

\bb{Nordtvedt}{1996}{Nordt}
Nordtvedt K., 1996, Class. Quantum Gravit., 13, A11

\bb{Pitjeva}{2010}{Pit}
Pitjeva E.V, 2010, EPM ephemerides and relativity. In:  Klioner S.A.,   Seidelmann P.K.,  Soffel M.H.,  (eds.)  Relativity in Fundamental Astronomy: Dynamics,
Reference Frames, and Data Analysis,
Proceedings IAU Symposium No. 261, Cambridge University Press, Cambridge, 2010, pp. 170-178.

\bb{Reynaud \& Jaekel}{2005}{Rey}	
Reynaud S., Jaekel M.-Th., 2005, I. J. Mod. Phys. A, 20,  2294

\bb{Roncoli}{2005}{luna}
Roncoli R. B., 2005, Lunar Constants and Models Document, JPL D-32296, Jet Propulsion Laboratory,
California Institute of Technology

\bb{Ruggiero}{2010}{Rug}
Ruggiero M. L., 2010, arXiv:1010.2114

\bb{Ruggiero \& Iorio}{2010}{RugIor}
Ruggiero M. L., Iorio L., 2010, Gen. Relativ. Gravit., 42, 2393

\bb{Sanders}{2006}{San}
Sanders R. H., 2006, Mon. Not. R. Astron. Soc., 370, 1519

\bb{Sch\"{a}fer \& Wex}{1993}{2pn2}
Sch\"{a}fer G.,  Wex N., 1993, Phys. Lett. A, 174, 196

{
\bb{Seidelmann}{1992}{seidel}
Seidelmann P. K. (ed.), 1992, Explanatory Supplement to the Astronomical Almanac,
University Science Books, Mill Valley, CA,
}


\bb{Soffel et al.}{1988}{Sof1}
Soffel M., Wirrer R., Schastok J., Ruder H., Schneider M., 1988, Celest. Mech. Dyn. Astron., 42, 81



\bibitem[\protect\citeauthoryear{Soffel}{1989}]{Soff}
Soffel M.H., 1989,  {Relativity in Astrometry, Celestial Mechanics and
Geodesy}, Springer Verlag, Berlin

\bb{Sotiriou \& Faraoni}{2010}{Sot}
Sotiriou T., Faraoni V., 2010, Rev. Mod. Phys., 82, 451

%\bb{Standish}{2008}{Sta}
%Standish E. M., Planetary and Lunar Ephemerides: testing alternate gravitational theories, 2008,
%In: Macias A.,  L\"{a}mmerzahl C.,  Camacho A., (eds.) AIP Conference Proceedings Volume 977. RECENT DEVELOPMENTS IN
%GRAVITATION AND COSMOLOGY: 3rd Mexican Meeting on Mathematical and Experimental
%Physics,  American Institute
%of Physics, p. 254.



\bb{Wex}{1995}{2pn3}
Wex N., 1995, Class. Quantum Gravit., 12, 983
{
\bb{Williams et al.}{2001}{JGR}
Williams J. G., Boggs D. H., Yoder C. F., Ratcliff J. T., Dickey J. O., 2001, J. Geophys. Res., 106, 27933
}

{
\bb{Williams \& Dickey}{2003}{hjk}
Williams J. G.,  Dickey J. O., 2003, Lunar Geophysics, Geodesy, and Dynamics. In: Noomen R., Klosko S., Noll C.,  Pearlman M. (eds.) Proceedings of the 13th International Workshop on Laser Ranging,  NASA/CP-2003-212248, pp. 75-86. http://cddis.nasa.gov/lw13/docs/papers/sci$\_$williams$\_$1m.pdf
}

\bb{Williams et al.}{2007}{Willi}
 Williams J. G.,  Turyshev S. G.,  Boggs D. H., 2007, Phys. Rev. Lett.,  98,  059002

\bb{Williams et al.}{2008}{de421b}
Williams J. G.,  Boggs D. H.,  Folkner W. M., 2008, DE421 Lunar Orbit, Physical Librations, and Surface Coordinates, JPL IOM 335-JW,DB,WF-20080314-001

\bb{Williams \& Boggs}{2009}{WiBo}
Williams J. G.,   Boggs D. H., 2009, Lunar Core and Mantle. What Does LLR See? In: Schilliak S. (eds.) Proceedings of the 16th International Workshop on Laser Ranging, {pp. 101-120}. http://cddis.gsfc.nasa.gov/lw16/docs/papers/sci$\_$1$\_$Williams$\_$p.pdf

\bb{Xu \& Siegel}{2008}{Xu}
Xu X.,  Siegel E. R., 2008,  arXiv:0806.3767

\end{thebibliography}
\end{document}